\documentclass[journal]{article}
\usepackage{amsfonts,amssymb,anysize}
\usepackage{times,mathptm}
\usepackage{amsmath,amsthm,graphicx,subfigure}
\marginsize{0.9in}{0.9in}{0.6in}{0.8in}

\newtheorem{example}{Example}

\usepackage[matrix,frame,arrow]{xy}
\usepackage{amsmath}

\newcommand{\ket}[1]{\left\vert{#1}\right\rangle}
\newcommand{\qw}[1][-1]{\ar @{-} [0,#1]}
\newcommand{\qwx}[1][-1]{\ar @{-} [#1,0]}

\newcommand{\gate}[1]{*{\xy *+<.6em>{#1};p\save+LU;+RU **\dir{-}\restore\save+RU;+RD **\dir{-}\restore\save+RD;+LD **\dir{-}\restore\POS+LD;+LU 
**\dir{-}\endxy} \qw}

\newcommand{\control}{*!<0em,.025em>-=-{\bullet}}
\newcommand{\controlo}{*-<.21em,.21em>{\xy *=<.59em>!<0em,-.02em>[o][F]{}\POS!C\endxy}}
\newcommand{\ctrl}[1]{\control \qwx[#1] \qw}
\newcommand{\ctrlo}[1]{\controlo \qwx[#1] \qw}
\newcommand{\targ}{*!<0em,.019em>=<.79em,.68em>{\xy {<0em,0em>*{} \ar @{ - } +<.4em,0em> \ar @{ - } -<.4em,0em> \ar @{ - } +<0em,.36em> \ar @{ - } 
-<0em,.36em>},<0em,-.019em>*+<.8em>\frm{o}\endxy} \qw}

\newcommand{\multigate}[2]{*+<1em,.9em>{\hphantom{#2}} \qw \POS[0,0].[#1,0];p !C *{#2},p \save+LU;+RU **\dir{-}\restore\save+RU;+RD 
**\dir{-}\restore\save+RD;+LD **\dir{-}\restore\save+LD;+LU **\dir{-}\restore}
\newcommand{\ghost}[1]{*+<1em,.9em>{\hphantom{#1}} \qw}

\newcommand{\lstick}[1]{*!R!<.5em,0em>=<0em>{#1}}

\newcommand{\Qcircuit}[1][0em]{\xymatrix @*[o] @*=<#1>}

\begin{document}

\title{Reversible Circuit Optimization via Leaving the Boolean Domain\thanks{
$\copyright$ 2011 IEEE.  Reprinted, with permission, from IEEE TCAD, 30(6):806--816, 2011. \newline
This material is posted here with permission of the IEEE.  Internal or
personal use of this material is permitted.  However, permission to
reprint/republish this material for advertising or promotional purposes or
for creating new collective works for resale or redistribution must be
obtained from the IEEE by writing to pubs-permissions@ieee.org. \newline
By choosing to view this document, you agree to all provisions of the
copyright laws protecting it.}}

\author{Dmitri~Maslov\thanks{D. Maslov is with the Institute for Quantum Computing, University of Waterloo,
Waterloo, ON, N2L 3G1, Canada.} and
		Mehdi~Saeedi\thanks{M. Saeedi is with the Computer Engineering Department, Amirkabir University of Technology, Tehran, Iran.}
		}


\maketitle

\begin{abstract}
For years, the quantum/reversible circuit community has been convinced that: {\em a)} the addition
of auxiliary qubits is instrumental in constructing a smaller quantum circuit; and,
{\em b)} the introduction of quantum gates inside reversible circuits may result in more efficient designs.
This paper presents a systematic approach to optimizing reversible (and quantum) circuits via
the introduction of auxiliary qubits and quantum gates inside circuit designs. This advances
our understanding of what may be achieved with {\em a)} and {\em b)}.
\end{abstract}



\section{Introduction}
Quantum computing \cite{bk:nc} is a computing paradigm studied for two major reasons:
\begin{itemize}
\item The associated complexity class, BQP, of the problems solvable by a quantum algorithm
in polynomial time, appears to be larger than the class P of problems solvable by a deterministic
Turing machine (in essence, a classical computer) in polynomial time. One of the best known examples
of a quantum algorithm yielding a complexity reduction when compared to the best known classical
algorithm includes the ability to find a discrete logarithm
over Abelian groups in polynomial time (this includes Shor's famous integer factorization algorithm as
a special case when the group considered is $\mathbb{Z}_m$). In particular, a discrete logarithm over an elliptic
curve group over $GF(2^m)$ can be found by a quantum circuit with $O(m^3)$ gates \cite{ar:pz}, whereas the best
classical algorithm requires a fully exponential $O(\sqrt{2}^m)$ search.
\item Quantum computing is physical, that is, quantum mechanics defines how
a quantum computation should be done.  With our current knowledge, it is perfectly feasible to
foresee hardware that directly realizes quantum algorithms, {\it i.e.}, a quantum computer.
It is generally perceived that challenges in realizing large-scale quantum computation are
technological, as opposed to a flaw in the formulation of quantum mechanics.
\end{itemize}

To have an efficient quantum computer means not only be able to derive favorable complexity
figures using big-$O$ notation and be able to control quantum mechanical systems with a high fidelity and
long coherence times, but also to have an efficient set of Computer Aided Design tools. This is similar to classical computation.
A Turing machine paradigm, coupled with high clock speed and no errors in switching, is not sufficient
for the development of the fast classical computers that we now have.
However, due to a great number of engineering solutions, including CAD, we are able to create very fast classical computers.

To the best of our knowledge, true reversible circuits are currently limited to the quantum technologies.
All other attempts to implement reversible logic are based on classical technologies, {\it e.g.}, CMOS,
and, internally, they are not reversible. For those latter internally irreversible technologies, it
may not be beneficial to consider reversible circuits, since reversibility is a restriction that
complicates circuit design\footnote{A reversible design is a combinational circuit, but not every combinational
circuit is necessarily reversible, moreover, most are not.}, but does not provide a speed-up or a lower power consumption/dissipation due
to the internal irreversibility of the underlying technology. In quantum computing, however, reversibility is out of
necessity (apart from the measurements that are frequently performed at the end of a quantum computation).

Reversible circuits are an important class of computations that needs to be performed efficiently for the
purpose of efficient quantum computation.  Indeed, multiple quantum algorithms contain arithmetic units
({\it e.g.}, adders, multipliers, exponentiation, comparators, quantum register shifts and permutations)
that are best viewed as reversible circuits; reversible circuits are indispensable for quantum error
correction.  Often, efficiency of the reversible implementation is the bottleneck of a quantum algorithm
({\it e.g.}, integer factoring and discrete logarithm \cite{bk:nc}) or a class of quantum circuits ({\it e.g.},
stabilizer circuits \cite{ar:ag}).

In this paper, we describe an algorithm that, in the presence of auxiliary qubits set to value $\ket{0}$,
rewrites a suitable reversible circuit into a functionally equivalent quantum circuit with a lower
implementation cost.  We envision that for all practical purposes, a reversible transformation is likely
a subroutine in a larger quantum algorithm.  When implemented in the circuit form, such a quantum algorithm
may benefit from extra auxiliary qubits carried along to optimize relevant quantum implementations and/or required for fault tolerance.  Those
auxiliary qubits may be available during the stages when a classical reversible transformation needs to be implemented,
and our algorithm intends to draw ancillae from this resource.  Our proposed optimization algorithm is best
employed at a high abstraction level,---before multiple control gates are decomposed into single- and two-qubit gates.

\section{Related work}
In existing literature, ignoring modifications, there are three basic algorithms for reversible circuit optimization.
\begin{itemize}
\item Template optimization \cite{ar:mmd}. Templates are circuit identities. They possess the property that a continuous
subcircuit cut from an identity circuit is functionally equivalent to a combination of the remaining gates.
A template application algorithm matches and moves as many gates as possible based on the description of a template.
It then replaces the gates with a different, but simpler circuit, as specified by the particular template being used.
\item A variation of peephole optimization \cite{ar:pspmh}. This algorithm optimizes a reversible circuit composed
with NOT, CNOT and Toffoli gates. The algorithm relies on a database storing optimal implementations of all 3-bit reversible
circuits and some small 4-bit implementations. It then finds a continuous subcircuit within a circuit to be simplified
such that gates in it operate on no more than 4 bits. Following this, it computes the functionality of this piece and replaces with an optimal
implementation when possible to find one. This algorithm is {\em not} limited to NOT, CNOT and Toffoli library, rather,
it relies heavily on the number of optimal implementations that could be accessed, and an efficient algorithm for finding
and/or transforming a target circuit into the one having a large continuous piece that allows simplification.
\item Resynthesis ({\em e.g.}, \cite{ar:mmd}). In its most general formulation, this is an approach where a subcircuit of a given
circuit is resynthesized, and if the result of such resynthesis is a preferred implementation, the replacement is
done. Peep-hole optimization is a type of such generic interpretation of the resyntheis. The authors of \cite{ar:mmd} used a heuristic
to perform resynthesis and did not limit the number of bits in a circuit to be resynthesized.
\end{itemize}

Recently, a BDD-based (Binary Decision Diagram-based) reversible logic synthesis algorithm was introduced \cite{co:wd}.
This algorithm employs ancillary bits to synthesize reversible circuits.  In principle, this synthesis algorithm could be
turned into a circuit optimization approach via employing it as a part of resynthesis.  However, this approach appears
to be inefficient due to the tendency of the synthesis algorithm to use both a larger number of qubits and a larger number
of gates than other reversible logic synthesis algorithms.

\section{Preliminaries} \label{sec:basic_concepts}
A qubit (quantum bit) is a mathematical object that represents the state of an elementary quantum mechanical system using its two basic 
states---$\ket{0}$, a low energy state, and $\ket{1}$, a high energy state.  Moreover, any such elementary single qubit quantum system may be described by 
a linear combination of its basic states, $\ket{\psi} = \alpha\ket{0}+\beta\ket{1}$,
where $\alpha$ and $\beta$ are complex numbers.

Upon measurement (computational basis measurement), the state collapses into one of the basis vectors, $\ket{0}$ or $\ket{1}$, with the probability of 
$|\alpha|^2$ and $|\beta|^2$, respectively (consequently, $|\alpha|^2+|\beta|^2=1$).  A quantum $n$-qubit system $\ket{\phi}$ is a tensor product of the 
individual single qubit states, $\ket{\phi} = \ket{\psi_1} \otimes \ket{\psi_2} \otimes ... \otimes \ket{\psi_n}$.  Furthermore, quantum mechanics 
prescribes that the evolution of a quantum $n$-qubit system is described by the multiplication of the state vector by a proper size unitary matrix $U$ (a 
matrix $U$ is called unitary if $UU^\dag=I$, where $U^\dag$ is the conjugate transpose of $U$ and $I$ is the identity matrix).  As such, the set of states 
of a quantum system forms a linear space.  A vector/state $\ket{\phi_\lambda}$ is called an eigenvector of an operator $U$ if $U\ket{\phi_\lambda} = 
\lambda\ket{\phi_\lambda}$ for some constant $\lambda$.  The constant $\lambda$ is called the eigenvalue of $U$ corresponding to the eigenvector 
$\ket{\phi_\lambda}$.

An $n$-qubit quantum gate performs a specific $2^n\times2^n$ unitary operation on the selected $n$ qubits it operates on in a specific period of time.  
Previously, various quantum gates with different functionalities have been described.  Among them, the CNOT (controlled NOT) acts on two qubits (control 
and target) where the state of the target qubit is inverted if the control qubit holds the value $\ket{1}$.  The matrix representation for the CNOT gate 
is:
\[
\left[ {\begin{array}{*{20}c}
   1 & 0 & 0 & 0  \\
   0 & 1 & 0 & 0  \\
   0 & 0 & 0 & 1  \\
   0 & 0 & 1 & 0  \\
\end{array}} \right]
\]

The Hadamard gate, $H$, maps the computational basis states as follows:
\[
\begin{array}{l}
 H\left| 0 \right\rangle  = \frac{1}{{\sqrt 2 }}(\left| 0 \right\rangle  + \left| 1 \right\rangle ) \\
 H\left| 1 \right\rangle  = \frac{1}{{\sqrt 2 }}(\left| 0 \right\rangle  - \left| 1 \right\rangle ) \\
 \end{array}
\]

The Hadamard gate has the following matrix representation:

\[
\frac{1}{{\sqrt 2 }}\left[ {\begin{array}{*{20}c}
   1 & 1  \\
   1 & { - 1}  \\
\end{array}} \right]
\]

The unitary transformation implemented by one or more gates acting on different qubits is calculated as the tensor product of their respective matrices 
(if no gate acts on a given qubit, the corresponding matrix is the identity matrix, $I$).  When two or more gates share a qubit they operate on, most 
often, they need to be applied sequentially.  For a set of $k$ gates $g_1$, $g_2$, ..., $g_k$ forming a quantum circuit $C$, the unitary calculated by $C$ 
is described by the matrix product $M_k M_{k-1} ... M_1$ where $M_i$ is the matrix of $i^{th}$ gate ($1\leq i \leq k$).

Given any unitary gate $U$ over $m$ qubits $\ket{x_1 x_2  \cdots \,x_m}$, a controlled-$U$ gate with $k$ control qubits $\ket{y_1 y_2  \cdots \,y_k}$ may 
be defined as an $(m+k)$-qubit gate that applies $U$ on $\ket{x_1 x_2  \cdots \,x_m }$ iff $\ket{y_1 y_2  \cdots \,y_k}$=$\ket{1}^{\otimes k}$ (we use 
$\ket{1}^{\otimes k}$ to denote the tensor product of $k$ qubits, each of which resides in the state $\ket{1}$).  For example, CNOT is the controlled-NOT 
with a single control, Toffoli gate is a NOT gate with two controls, and Fredkin gate is the controlled-SWAP (a SWAP gate maps $\ket{ab}$ into $\ket{ba}$) 
with a single control.

For a circuit $C_U$ implementing a unitary $U$, it is possible to implement a circuit for the controlled-$U$ operation by replacing every gate $G$ in 
$C_U$ by a controlled gate controlled-$G$.  It is often useful to consider unitary gates with control qubits set to value zero.  In circuit diagrams, 
$\circ$ is used to indicate conditioning on the qubit being set to value zero (negative control), while $\bullet$ is used for conditioning on the qubit 
being set to value one (positive control).

In this paper, we consider reversible circuits.  A reversible gate/operation is a $0-1$ unitary, and reversible circuits are those composed with 
reversible gates.  A \emph{multiple control Toffoli gate} C$^m$NOT $(x_1, x_2, \cdots, x_{m+1})$ passes the first $m$ qubits unchanged.  These qubits are 
referred to as \emph{controls}.  This gate flips the value of $(m+1)^{th}$ qubit if and only if the control lines are all one (positive controls).  
Therefore, action of the multiple control Toffoli gate may be defined as follows: $x_{i(out)}=x_i (i<m+1), x_{m+1(out)}=x_1 x_2 \cdots x_m \oplus 
x_{m+1}$.  Negative controls may be applied similarly.  For $m=0$, $m=1$, and $m=2$ the gates are called NOT, CNOT, and Toffoli, respectively.

It has been shown that there are a number of problems that may be solved more efficiently by a quantum algorithm, as opposed to the best known classical 
algorithm.  One such algorithm is the Deutsch-Jozsa algorithm \cite{ar:dj}.  To illustrate this algorithm, let $f : \{0, 1\}\rightarrow \{0, 1\}$ be a 
single-input single-output Boolean function.  Note that there are only four possible single-input single-output functions, namely, $f_1(x)=0$, $f_2(x)=1$, 
$f_3(x)=x$, $f_4(x)=\bar{x}$.  We can easily verify that $f_1$ and $f_2$ are constant, and $f_3$ and $f_4$ are balanced (meaning the number of ones in the 
output vector is equal to the number of zeroes).  Imagine we have a black box implementing function $f$, but we do not know which kind it is---constant or 
balanced.  The goal is to classify this function, and one is allowed to make queries to the black box.  With classical resources, we need to evaluate $f$ 
twice to tell, with certainty, if $f$ is constant or balanced.  However, there exists a quantum algorithm, known as Deutsch-Jozsa algorithm, that performs 
this task with a single query to $f$.  Figure \ref{fig:Deutsch} shows the quantum circuit implementing the Deutsch-Jozsa algorithm where $U_f: \ket{x,y} 
\mapsto \ket{x,y \oplus f(x)}$.  The quantum state (Figure \ref{fig:Deutsch}) evolves as follows:

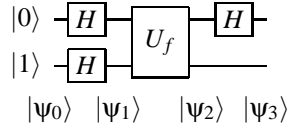
\begin{figure}
\[
\Qcircuit @C=0.5em @R=0.5em {
\lstick{\ket{0}}   & \gate{H} 				& \qw				& \multigate{1}{U_f} & \qw				& \gate{H} 		& \qw	\\
\lstick{\ket{1}}   & \gate{H} 				& \qw				& \ghost{U_f} 			& \qw				& \qw				& \qw	\\
						 &								&					&							&					&					&		 
\\
						 & \lstick{\ket{\psi_0}}& \ket{\psi_1} & 							& \ket{\psi_2}	& 					& \ket{\psi_3}\\
}
\]
\centering
\caption{Quantum circuit implementing the Deutsch algorithm.}
\label{fig:Deutsch}
\end{figure}

\begin{figure}
    \centering
    \subfigure[\label{fig:Deutschf1}]{
        \Qcircuit @C=1.5em @R=2.2em @!R{
		&	\qw	&	\qw&	\\				
		&	\qw	&	\qw&	\\		
		&			&		&	\\
		}

    }
    \subfigure[\label{fig:Deutschf2}]{
        \Qcircuit @C=1.5em @R=1.5em @!R{
		&		\qw	&	\qw	\\				
		&		\targ	&	\qw	\\		
		&				&		\\	
		}

    }
    \subfigure[\label{fig:Deutschf3}]{
        \Qcircuit @C=1.5em @R=1.5em @!R{
		&		\ctrl{1} &	\qw	\\				
		&		\targ		&	\qw	\\		
		&					&			\\	
		}

   }
    \subfigure[\label{fig:Deutschf4}]{
        \Qcircuit @C=1.5em @R=1.5em @!R{
		&		\ctrlo{1}   &	\qw\\				
		&		\targ		&	\qw	\\	
		&					&	     \\	
	 	}

   }
    \caption{Four possible Deutsch-Jozsa oracles for a single-input function: (a) $f(x)=0$, (b) $f(x)=1$, (c) $f(x)=x$, (d) $f(x)=\bar{x}$.} 
\label{fig:Deutschdetail}
\end{figure}
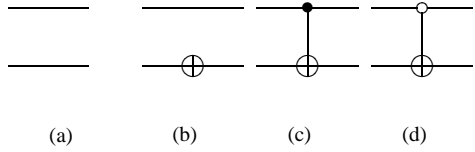

\[
\ket{\psi _0} = \ket{01}
\]

\[
\ket{\psi _1 } = \left[ {\frac{{\ket{0}  + \ket{1}}}{{\sqrt 2 }}} \right]\otimes\left[ {\frac{{\ket{0}  - \ket{1}}}{{\sqrt 2 }}} \right]
\]

\[
\ket{\psi _2 } = \left\{ \begin{array}{l}
  \pm \left[{\frac{{\ket{0} + \ket{1}}}{{\sqrt 2 }}} \right]\otimes\left[ {\frac{{\ket{0} - \ket{1}}}{{\sqrt 2 }}} \right]\,\,\,\,\,\,\,{\rm f(0) = f(1)} 
\\
  \pm \left[{\frac{{\ket{0} - \ket{1}}}{{\sqrt 2 }}} \right]\otimes\left[ {\frac{{\ket{0} - \ket{1}}}{{\sqrt 2 }}} \right]\,\,\,\,\,\,\,\,{\rm f(0)} \ne 
{\rm f(1)} \\
 \end{array} \right.
\]

\[
\ket{\psi _3 } = \pm \ket{f(0) \oplus f(1)} \otimes \left[{\frac{{\ket{0} - \ket{1}}}{{\sqrt 2 }}} \right].\,
\]

A measurement of the first qubit at the end of the circuit computes the value $f(0) \oplus f(1)$, which determines whether the function is constant or 
balanced.  Implementations of the $U_f$ for all four possible single-input functions $f$ are shown in Figure \ref{fig:Deutschdetail}.

\section{Problem formulation}

The circuit optimization algorithms discussed in the previous section are efficient, however, there is evidence that they will not
be able to discover all possible circuit simplifications.  In particular, it is generally believed that the addition of a number
of auxiliary bits may be instrumental in constructing a simpler circuit.

A classical example is the implementation of the $n$-bit
multiple control Toffoli gate \cite{ar:bbcd}.  Without any additional qubits, this gate may be implemented by a circuit requiring
$\Theta(n^2)$ two-qubit gates.  With the addition of a single qubit (and $n \geq 6$), the $n$-bit
multiple control Toffoli gate may be simulated by a circuit requiring a linear number of Toffoli gates, $8n+Const$, and as such,
a linear number of two-qubit gates.  With the addition of $(n-3)$ auxiliary bits (and $n \geq 4$), a more efficient implementation
requiring $4n+Const$ Toffoli gates is known.  Finally, if these $(n-3)$ auxiliary bits are set to value $\ket{00\ldots 0}$,
an even more efficient simulation requiring only $2n+Const$ Toffoli gates becomes available.  This is a clear indication that the addition of auxiliary
bits may be helpful in designing more efficient circuits.  However, at this point, no efficient methodology for automatic reversible circuit
simplification employing auxiliary bits has been suggested.  This paper presents such an algorithm.

When a Boolean function needs to be implemented in the circuit form, such a circuit may be composed solely of reversible gates or
it may be such a quantum circuit that, via leaving the Boolean domain, is capable of computing the desired function faster than any known classical
reversible circuit.  It is fair to say that a major goal of quantum computing as an area is to find as many problem-solution pairs such
that leaving the Boolean domain results in shorter computation.  An example of such a situation has been illustrated in the previous section by the 
Deutch-Jozsa algorithm.  This is a clear indication that significant speedups are possible via computing outside the Boolean
domain.  In this paper, we discuss an algorithm that rewrites a reversible circuit into a quantum circuit with a lower implementation
cost; for some circuits, it appears essential to have the ability to leave the Boolean domain to achieve simplification.
We illustrate performance by testing our algorithm on a set of benchmark functions.

In the remainder of the paper we assume a reversible circuit and a linear number of auxiliary bits prepared in the state
$\ket{00\ldots 0}$ are given as the input.  In other words, we are given a transformation $\ket{x}\ket{00...0} \mapsto RC\ket{x}\ket{00...0}$,
where it is guaranteed that $RC$ is a reversible circuit and $\ket{00...0}$ is not used in the computation.
Our goal is to rewrite this circuit into a quantum circuit that computes the same transformation
with lower implementation cost.  We do not assign a separate cost to the auxiliary qubits we use, but strictly limit
their quantity by the number of primary inputs in the reversible transformation, since reversible circuits
will most likely be used as subroutines in a larger quantum algorithm, whose implementation may require
extra ancillae to be available for error correction and to optimize implementation of different parts
of the quantum algorithm.  In other words, those auxiliary qubits may already be available.  If those ancillary qubits
are unavailable, or the additional cost associated with introducing them is too high, our proposed algorithm and its
implementation need to be updated.

\section{Algorithm}
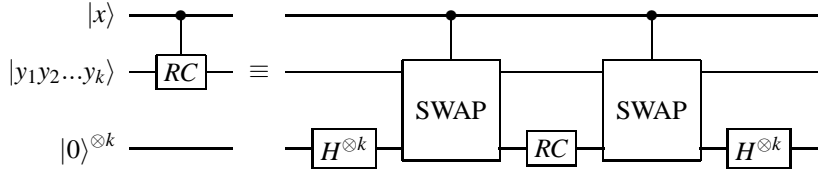
\begin{figure*}
\[
\Qcircuit @C=1em @R=1.5em {
\lstick{\ket{x}}              & \ctrl{1}  &\qw& \;     &&\qw                  & \ctrl{1}            & \qw      & \ctrl{1}            & \qw & \qw \\
\lstick{\ket{y_1y_2...y_k}}   & \gate{RC} &\qw& \equiv &&\qw      & \multigate{1}{\text{SWAP}} & \qw      & \multigate{1}{\text{SWAP}} & \qw & \qw \\
\lstick{\ket{0}^{\otimes k}}  & \qw       &\qw& \;     &&\gate{H^{\otimes k}} & \ghost{SWAP}        & \gate{RC}& \ghost{SWAP}        & \gate{{H}^{\otimes 
k}} & \qw \\
}
\]
\centering
\caption{Circuit equivalence. Note that it also holds if the control $\ket{x}$ is negative, when $\ket{x}$ is both
a single qubit or a quantum register, and a combination of these two (consistent positive and negative controls in
a register).}
\label{fig:main}
\end{figure*}

The idea behind our algorithm is best illustrated by the circuit equivalence shown in Figure \ref{fig:main}.

We first show correctness of this circuit equivalence.  The circuit on the left computes transformation
$\ket{1}\ket{y_1y_2...y_k}$ $\ket{00...0} \mapsto \ket{1}RC(\ket{y_1y_2...y_k})\ket{00...0}$ if the value of the
control variable $x$ is $1$ and otherwise,
$\ket{0}\ket{y_1y_2...y_k}\ket{00...0} \mapsto \ket{0}\ket{y_1y_2...y_k}\ket{00...0}$,
the identity function. The circuit on the right is composed of five stages/gates. The aggregate transformation
it computes for $x=1$ is (subject to normalization)
\begin{eqnarray*}
\ket{1}\ket{y_1y_2...y_k}\ket{00...0} \mapsto
\ket{1}\ket{y_1y_2...y_k}\sum_{i=0}^{2^k-1}\ket{i} \mapsto \\
\ket{1}\sum_{i=0}^{2^k-1}\ket{i}\ket{y_1y_2...y_k} \mapsto
\ket{1}\sum_{i=0}^{2^k-1}\ket{i}RC(\ket{y_1y_2...y_k}) \mapsto \\
\ket{1}RC(\ket{y_1y_2...y_k})\sum_{i=0}^{2^k-1}\ket{i} \mapsto
\ket{1}RC(\ket{y_1y_2...y_k})\ket{00...0},
\end{eqnarray*}
{\it i.e.}, it matches the computation performed on the left hand side. For value $x=0$ the transformation computed
is (subject to normalization)
\begin{eqnarray*}
\ket{0}\ket{y_1y_2...y_k}\ket{00...0} \mapsto \\
\ket{0}\ket{y_1y_2...y_k}\sum_{i=0}^{2^k-1}\ket{i} \mapsto
\ket{0}\ket{y_1y_2...y_k}\sum_{i=0}^{2^k-1}\ket{i} \mapsto \\
\ket{0}\ket{y_1y_2...y_k}\sum_{i=0}^{2^k-1}RC(\ket{i}) = \ket{0}\ket{y_1y_2...y_k}\sum_{i=0}^{2^k-1}\ket{i} \mapsto  \\
\ket{0}\ket{y_1y_2...y_k}\sum_{i=0}^{2^k-1}\ket{i} \mapsto
\ket{0}\ket{y_1y_2...y_k}\ket{00...0},
\end{eqnarray*}
{\it i.e.}, the identity, and thus matches the result of the computation in the circuit on the left. In the above,
the equality holds because the domain of any reversible transformation is the same set as its codomain. As such, an equal
weight superposition of all elements of the domain remains invariant under any reversible Boolean transformation.
In other words, since $H^{\otimes k}\ket{00...0}$ is an eigenvector of any $0-1$ unitary matrix $RC$ with
eigenvalue $1$, application of $RC$ to this eigenvector does nothing.  As a result, $RC$ may be applied uncontrollably
as long as we can control {\em what} it is being applied to---the desired vector or a dummy eigenvector, as opposed
to {\em how} ({\it i.e.}, in this case, controlled).
Furthermore, this identity is inspired by the generic construction of Kitaev \cite{arxiv:k}.

What makes this circuit identity practical for circuit simplification is a combination of:
the relative hardness of implementing many multiple control gates, frequent use of large
controlled blocks in circuit designs, the ease of the preparation
of the eigenvector $\sum_{i=0}^{2^k-1}\ket{i}$ (one layer of Hadamard gates), and
reusability of ancillae in the sense that Hadamards do not need to be uncomputed if the circuit
identity is to be applied once more to a different part of the circuit being simplified.

A layer of Hadamard gates is an eigenvector of any transformation computed by a reversible circuit, and as such it is
universally applicable in the above construction. However, if $RC$ has a fixed point $i$
such that $RC(i)=i$, rather than using Hadamards, one could ``hard code'' the value $i$ by applying
NOT gates at positions where binary expansion of $i$ is 1. The upside is that the number
of NOT gates that need to be applied does not exceed $k$, and generally their number is less than $k$.
Thus, the number of NOTs that need to be applied is expected to be less than the number of
Hadamards in the generic construction. This is, however, only a minor improvement due to the
relative ease of implementing NOT and Hadamard gates, and a small (at most, $2k$) number of those
required. The downside is that in sequential application of the circuit equivalence in
Figure \ref{fig:main}, the fixed point needs to be recoded for every new $RC$.

For reversible functions of $k$ bits, the number of those with no fixed points
is approximately $\frac{k!}{e} \approx .368k!$, where $e:=\lim_{n \rightarrow \infty}{(1+\frac{1}{n})^n} \approx 2.71828...$
\cite{wiki:perm}. To use the circuit equivalence in this case requires the use of Hadamard
gates. The number of Hadamard gates may be reduced to $s \leq k$ if the reversible circuit $RC$ is such that it fixes a Boolean cube
(meaning $RC(i) \in C$ for every $i \in C$, where $C$ is the Boolean cube) of size
$s$. For example, if the fixed cube is of the form ${--01-0}$, auxiliary qubits must be prepared
as follows: $H\otimes H\otimes I \otimes NOT\otimes H\otimes I \ket{000000}$ for the circuit
identity to work. In other words, for every variable changing its value, it requires
application of the Hadamard gate, for every variable taking the value $1$, application
of the NOT gate is required, and for every variable taking the value $0$, no gate is required.
An example of a reversible function requiring all $k$ Hadamards is the cycle shift
$(1,2,...,2^k-1,0)$, or any cycle of maximal length.
For all other permutations---those with at least one fixed point,
of which there are approximately $k!(1-\frac{1}{e}) \approx .632k!$,
we can find a proper set of NOT gates to use the circuit equivalence without any Hadamard gates.

Based on the above identity, the proposed reversible circuit optimization algorithm works as follows.
\begin{enumerate}
\item Prepare ancillae via applying a layer of Hadamard gates ($k-1$ bits suffices for any $k$-bit
circuit to be simplified).
\item Find sets of all possible adjacent gates sharing at least one common control.
\item Evaluate all sets of adjacent gates to find the one set that reduces the total cost more. When
such a set is found, apply the circuit identity shown in Figure \ref{fig:main}:
	\begin{enumerate}
	\item If we are dealing with the single shared control, apply the identity.
	\item If we dealing with a shared multicontrol, dedicate one of ancillary qubits as collecting the
	product defined by the shared multicontrol, and use this qubit to control the application of
	Fredkin gates. One has to be careful to make sure the chosen qubit has a correct combination of
	Hadamard gates on it, {\it i.e.}, an even number of Hadamards to achieve a Boolean value and store value
	of the control product, and an odd number if this bit is used for implementation of an
	uncontrolled transformation.
	\end{enumerate}
\item Update the remaining sets of adjacent gates to exclude all sets that intersect with the sets
already processed at step 3. If no sets remain, continue to the next step; otherwise, go to 3.
\item Calculate the number of auxiliary qubits we actually need in this process. Upper bound is $k-1$
for a $k$-bit circuit, but we can often do better than that due to the use of multicontrol
and tracking how many qubits the selected gates sharing a control operate on.  Also, since
there is a chance that all largest controlled gates in the circuit before simplification are
factored, we may need fewer extra qubits for an efficient implementation of the multiple
controlled gates.
\end{enumerate}
We have implemented this algorithm in C++, and report benchmark results in Section \ref{sec:pr}.

The above algorithm is very naive, and may be improved with the following modifications.
\begin{itemize}
\item Find more efficient ways to identify and process sets of gates sharing common controls.
Since our basic algorithm is greedy, there likely are better approaches than
finding all and picking the best found.
\item Find the simplest combination of NOT and Hadamard gates, as opposed to
using a layer of all $k$ Hadamard gates. If Hadamard gates need to be avoided at all
cost, $RC$ may be complemented by a minimal circuit $M$, followed by $M^{-1}$ such
that $RC \circ M$ has a fixed point. Then, controlled-$RC \circ M$ may be implemented
with the circuit identity, and $M^{-1}$ is not used in it. It is not clear
if leaving the Boolean domain is so unwelcome as to for this procedure to become efficient.
However, this gives birth to a new reversible circuit simplification approach
based solely within the Boolean domain.
\item Find better algorithms for collecting gates sharing common controls, {\it e.g.}, by finding
more efficient algorithms to move gates, as some non-commuting gates may be commuted through
a block of gates.

\begin{figure}[h]
\centering
\includegraphics[height=22mm]{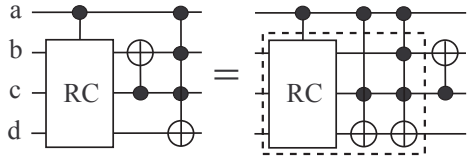}
\caption{Moving $TOF(a,b,c,d)$ to the left past $CNOT(c,b)$ via introducing $TOF(a,c,d)$.}
\label{fig:moveitsmart}
\end{figure}

More interestingly, consider the left circuit in Figure \ref{fig:moveitsmart}. It may be
rewritten in an equivalent form, as illustrated on the right in Figure \ref{fig:moveitsmart}.
At first glance, it may seem that the circuit on the right is more complex, since
it contains an extra gate, $TOF(a,c,d)$. However, as indicated by the dashed line,
$TOF(a,c,d)$ and $TOF(a,b,c,d)$ may now be merged into $RC$ and implemented
using the identity in Figure \ref{fig:main}. This was not possible before the transformation,
since gate $CNOT(c,b)$ was blocking $TOF(a,b,c,d)$ from joining the $RC$. The result
of this transformation is the effective ability to implement a multiple control Toffoli gate
with three controls for the cost of a Toffoli gate (with two controls) and a CNOT.
The latter is most likely more efficient.
\item Iterate our basic algorithm, {\it i.e.}, look for subcircuits sharing a common control
within subcircuits whose shared controls have been factored out.
\item Find other instances where the introduction of quantum gates helps optimize an implementation.
\end{itemize}

Efficiency of any such modification is highly dependent on the relation between costs
of NOT, Hadamard, CNOT, SWAP, Toffoli, Fredkin gates, {\it etc.}, as well as their
multiple control versions including those with negative controls, and the minimization criteria ({\it e.g.},
gate count {\it vs.} circuit depth {\it vs.} number of qubits {\it vs.} certain desirable fault tolerance properties, {\it etc.}).
In Section \ref{sec:pr} we consider the performance of a basic implementation of our algorithm, and
count the number of two-qubit gates used before and after simplification. This illustrates the efficiency
of our algorithm in the most generic scenario.
We conclude this section by illustrating how this algorithm works with two examples.


\begin{figure*}[t]
\centering
\includegraphics[height=90mm]{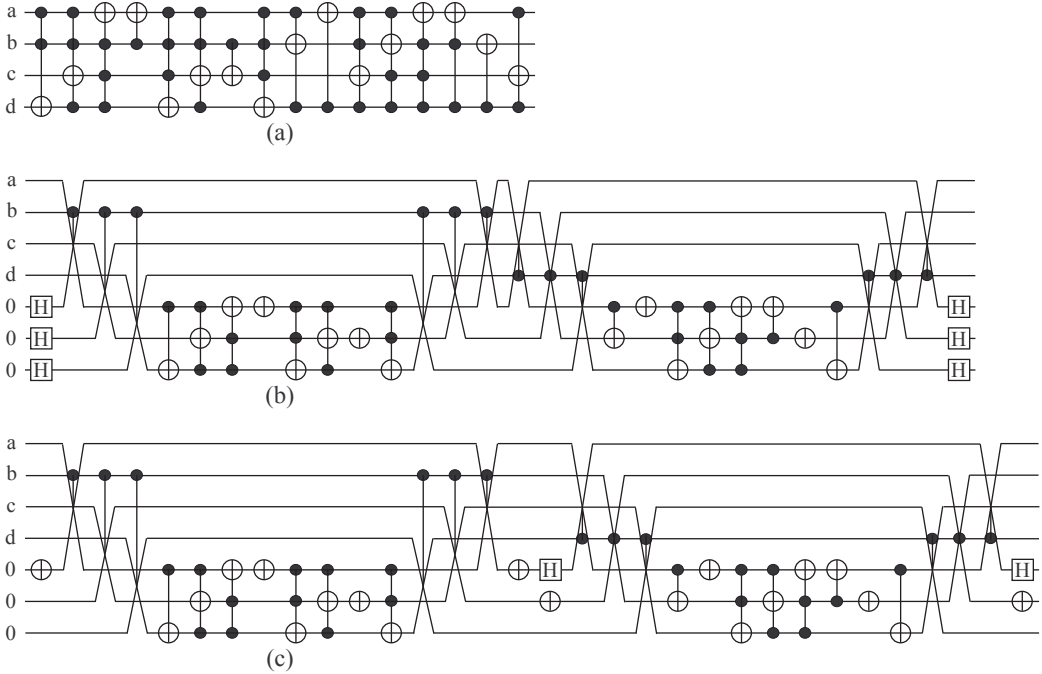}
\caption{Simplifying an example circuit (a): (b) using the algorithms introduced; (c) minimizing the number of Hadamard gates used.}
\label{fig:example}
\end{figure*}

\begin{example}
Illustrated in Figure \ref{fig:example}(a) is a circuit that we simplify using the suggested approach.
The initial circuit contains 4 two-qubit gates, 4 3-qubit gates and 8 4-qubit gates. Using a single number cost
estimation introduced in the next section, this circuit requires 144 two-qubit gates.
The algorithm, as described, finds two subcircuits sharing control variable $b$ in the first and control
variable $d$ in the second. Those subcircuits are implemented on a separate 3-qubit register
and copied in when required, as shown in Figure \ref{fig:example}(b).
The new circuit contains 10 single-qubit gates, 4 two-qubit gates, and
20 3-qubits gates. In other words, its implementation requires 104 two-qubit gates.
To construct the bottom circuit illustrated in Figure \ref{fig:example}, one needs to
notice that $\{a=1,c=0,d=0\}$ is a fixed point of the function
computed by the first subcircuit (after control $b$ is factored out),
and the second subcircuit (once control $d$ is cut) fixes the Boolean cube $\{a=variable,b=1,c=0\}$.
\end{example}

\begin{figure*}[t]
\centering
\includegraphics[height=100mm]{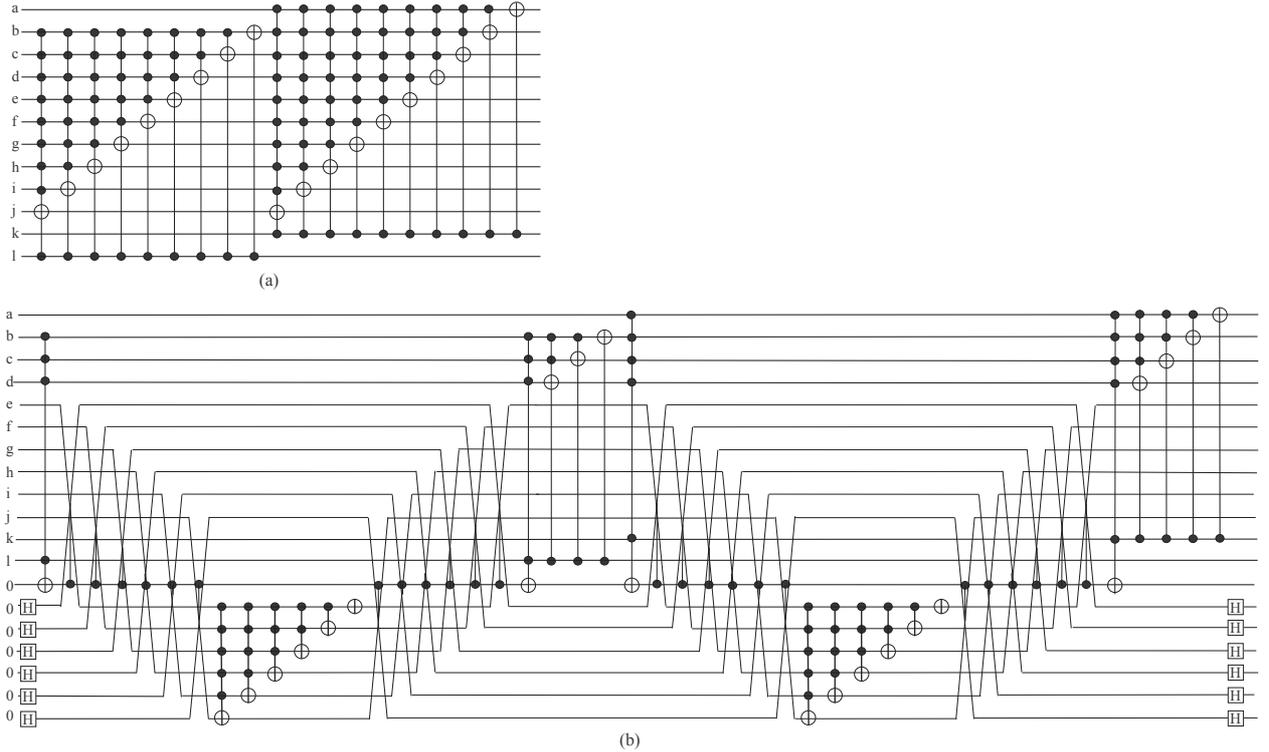}
\caption{Simplifying the cycle10\_2 circuit \cite{www:rlsb}, (a) the original circuit with cost 727, (b) the improved circuit with cost 469.}
\label{fig:examplecycle102}
\end{figure*}

\begin{example}
As an example with shared multicontrols consider the circuit cycle10\_2 \cite{www:rlsb} shown in Figure \ref{fig:examplecycle102}(a).  As can be seen, the 
circuit has several gates with shared common controls.  According to the proposed algorithm, in this case, one of the auxiliary qubits should be used to 
collect the product defined by the shared multicontrol and to control the application of Fredkin gates.  To find the appropriate set of common controls, 
the cost of the circuit before and after the optimization should be examined.  Using the single number cost estimation introduced in the next section, if 
the first 6 gates in Figure \ref{fig:examplecycle102}(a) are considered as a subcircuit with three shared common controls, the resulting implementation 
cost will be maximally improved.  Similarly, another subcircuit with 6 gates sharing 4 common controls can be recognized and optimized.  The resulting 
improved circuit is shown in Figure \ref{fig:examplecycle102}(b).  Altogether, the cost of the original circuit is improved by about 35\% (727 {\it vs.} 
469).
\end{example}

\section{Performance and Results} \label{sec:pr}
Before we can test the performance of the introduced approach, it is important to establish
a metric to define the implementation cost of a circuit before and after simplification.

\subsection{Circuit Cost} \label{ss:cc}

With our approach, we allow auxiliary qubits, which directly affects the cost of multiple control gates.  Further, we allow those qubits to carry value 
$\ket{00...0}$, which also affects how efficiently one is able to implement multiple control gates.  Due to these changes from convention, most common 
circuit cost metrics used, {\it e.g.}, \cite{www:rlsb, ar:mmd, ar:pspmh}, cannot be applied.  As a result, it is necessary to revisit the circuit cost 
metric.

Particulars of the definition of the cost metric largely affect practical efficiency.  We thus consider a
very generic definition of the circuit cost, and suggest that it is re-evaluated in the
scenario when circuit costs may be calculated with a better accuracy, and our algorithm/implementation
is updated correspondingly.

We will evaluate circuit implementation cost via estimating the number of two-qubit gates required to
implement it.

We ignore single-qubit gates partially because they may be merged into two-qubit gates (for instance,
in an Ising Hamiltonian\footnote{For the purpose of this paper, it suffices to state that an Ising Hamiltonian 
is such that the two-qubit interaction terms are described by the formula $\sum_{i<j}J_{ij}\sigma_z^i\sigma_z^j$,
where $\sigma_z^i$ is the Pauli-Z matrix acting on qubit $i$, and each $J_{ij}$ is a constant.} 
\cite{bk:nc} CNOT$(a,b)$NOT$(b)$ may be implemented as efficiently as CNOT$(a,b)$, and
$R^b_y(\pi/2)$CNOT$(a,b)$ is more efficient than CNOT$(a,b)$ on it own), and
partially because they are relatively easy to implement as compared to the two-qubit gates.

Efficiency of the implementation of the two-qubit gates depends on the Hamiltonian describing
the physical system being used.  For instance, in an Ising Hamiltonian, and up to single-qubit gates,
CNOT is equivalent to a single use of the two-qubit interaction term, $ZZ$.  With Ising Hamiltionian,
SWAP requires three uses of the interaction term, which is a maximum for the number of times
an interaction term needs to be used to implement any two-qubit gate in any Hamiltonian \cite{ar:zvs}.
However, if the underlying Hamiltonian is Heisenberg/exchange type \cite{bk:nc, ar:wab}, SWAP is implemented with a single
use of the two-qubit interaction, $XX+YY+ZZ$, and CNOT is notably more complex than SWAP.  For the sake of simplicity, we count
all two-qubit gates as having the same cost, and assign this cost a value of 1.

Efficient decomposition
of the Toffoli and Fredkin gates into a sequence of two-qubit gates largely depends on what
physical system is being used.  A Toffoli gate may be implemented up to a global phase using at most
3 two-qubit gates (all CNOTs, plus some single-qubit gates) or exactly using 5 two-qubit gates \cite{bk:nc}.
Other more efficient implementations are possible in very specific cases, {\it e.g.},
3 two-qubit gates suffice when the output is computed onto a qutrit (as opposed to a qubit) \cite{ar:rrg}.
The best known implementation of the Fredkin gate requires 3 pulses, each of which is a two-qubit gate \cite{ar:fjm}.
Finally, since when conjugated from left and right by a proper CNOT gate, a Toffoli gate becomes a Fredkin gate,
and a Fredkin gate becomes a Toffoli gate, their two-qubit gate implementation costs are within $\pm 2$ of each
other.  For the purpose of this paper, we will assign a cost of 5 to both Toffoli and Fredkin gates
(minimal two-qubit implementation cost reported in the literature plus 2).  Any other number between 3 and 7
would have been reasonable too.

\begin{figure}[h]
\centering
\includegraphics[height=30mm]{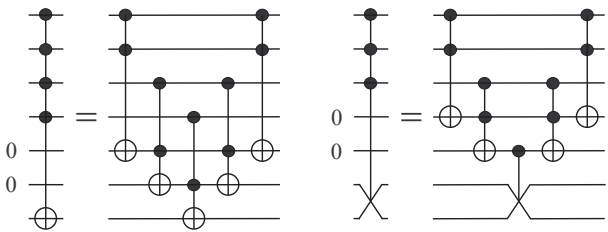}
\caption{Implementation of multiple control Toffoli and Fredkin gates \cite{bk:nc}.}
\label{fig:tof-fre}
\end{figure}

Multiple control Toffoli and multiple control Fredkin gates may be simulated such as shown
in Figure \ref{fig:tof-fre}. As such, both $n$-qubit Toffoli and $n$-qubit Fredkin gates
($n \geq 3$) require $2n-5$ 3-qubit Toffoli and Fredkin gates each, which translates
into $10n-25$ two-qubit gates. Since we ignore single-qubit gates, multiple control Toffoli
and Fredkin gates with negated controls have the same cost as their alternatives with positive
controls.

\subsection{Benchmarks}
We have experimented with those MCNC benchmarks we were able to find, and those circuits available at \cite{www:rlsb}.
Reversible circuits for some MCNC benchmarks were reported in \cite{ws:mp}
(top third of Table \ref{tab:bench}), and for the most popular that were not explicitly reported in \cite{ws:mp}
(middle third of Table \ref{tab:bench}), we used EXORCISM-4 \cite{mish} to synthesize them.  Finally,
we included circuits from \cite{www:rlsb} (bottom third of Table \ref{tab:bench}).
To save space, we report simplification of only those circuits that were the best reported
in the literature at the time of this writing; {\it e.g.}, \cite{ws:mp} reports a circuit for
function $rd73$, however, a better circuit exploiting the fact that this function is symmetric
is known \cite{www:rlsb}. Similarly, we found a number of simplifications in the $hwb$ type circuits,
however, we do not report those since efficient circuits for this family of functions have been found \cite{www:rlsb}.
Our approach is most efficient when applied to the circuits with a large proportion of multiple controlled gates.
Consequently, we did not find simplifications in the circuits dominated by small gates.

Table \ref{tab:bench} reports the results.  The first column lists a circuit index number that is introduced to be used in Table 
\ref{tab:bench2} as a reference.  The next two columns describe the original benchmark function,
including its name ({\bf name}), and the number of inputs and outputs ({\bf I/O}).  
The next four columns describe the best known
reversible circuit implementations. The first column, {\bf \# qubits}, lists the number of actual qubits
used, assuming every multiple control Toffoli gate is implemented most efficiently using a number of
auxiliary qubits (Figure \ref{fig:tof-fre}). This is why this number is higher than the sum of inputs
and outputs for irreversible specifications, and the number of inputs/outputs for reversible specifications.
The next column, {\bf \# rev. gates}, lists the number of multiple control reversible gates used,
{\bf cost} shows the cost, as defined in Subsection \ref{ss:cc}, {\it i.e.}, the number of two-qubit gates required,
and {\bf source} shows where or how this circuit may be obtained. The following four columns
summarize our simplification results, including the number of actual qubits required in the simplified circuits,
the number of gates in the new designs ({\bf \# quant. gates} - {\bf \# rev. gates} = number of
Hadamard and Fredkin gates our algorithm introduces), and cost of the simplified circuits.  Finally,
the last column, {\bf \% improvement}, shows the percentage of the reduction in cost as a result of the
application of our algorithm.

Table \ref{tab:bench2} presents the distribution of the number of gates in the circuits $b$efore and $a$fter simplification.
Each circuit is marked with $ix$, where $i$ is the circuit index number taken from Table \ref{tab:bench},
and $x$ takes values $b$ and $a$, to distinguish circuits $b$efore and $a$fter the simplification.  Columns report
the gate counts used in the corresponding circuit designs.  The columns are marked to represent the gate types used: 
NOT (T1), CNOT (T2), Toffoli (T3), ..., Toffoli-21 (T21), Fredkin (F3), and Hadamard (H).   

Most circuits were analyzed and simplified almost instantly.  The runtime depends primarily on the number
of gates, and the complexity of combinations of shared control configurations.  The longest computation
took 323 seconds (user time) to analyze circuit for $apex4$ function with 5131 gates.  It took 25 seconds
for the second largest circuit with 1917 gates, implementing the benchmark function $seq$.
We did not attempt to optimize our implementation.

\begin{table}
\begin{footnotesize}
\begin{center}
\caption{Benchmark results.  The actual circuit designs are available at \texttt{http://www.iqc.ca/\~\;dmaslov/rev2quant/},
and may be viewed with RCViewer+ available at \texttt{http://ceit.aut.ac.ir/QDA/RCV.htm}.}
\begin{tabular}{|l|c|c|c|c|c|c|c|c|c|c|}\hline
&\multicolumn{2}{|c|}{Function} & \multicolumn{4}{|c|}{Best Known Implementation} & \multicolumn{3}{|c|}{After Optimization} & \% \\
ckt\# &name & I/O   & \# qubits & \# rev. gates & cost   & source       & \# qubits & \# quant. gates & cost  & improvement \\ \hline
1 & 5xp1 & 7/10  & 22        & 61                  & 1177   & \cite{ws:mp} & 32        & 141      & 927   & 21.24\% \\
2 & add6 & 12/7  & 24        & 188                 & 6120   & \cite{ws:mp} & 40        & 330      & 3551  & 41.98\% \\
3 & b12  & 15/9  & 30        & 43                  & 1199   & \cite{ws:mp} & 41        & 113      & 831   & 30.17\% \\
4 & clip & 9/5   & 21        & 120                 & 5412   & \cite{ws:mp} & 31        & 296      & 2924  & 45.97\% \\
5 & in7  & 26/10 & 51        & 70                  & 4228   & \cite{ws:mp} & 65        & 190      & 2287  & 45.91\% \\
6 & life & 9/1   & 17        & 50                  & 2480   & \cite{ws:mp} & 24        & 152      & 1870  & 24.6\% \\
7 & ryy6 & 16/1  & 30        & 40                  & 2686   & \cite{ws:mp} & 40        & 134      & 1737  & 35.33\% \\
8 & sao2 & 10/4  & 22        & 58                  & 3972   & \cite{ws:mp} & 30        & 164      & 1806  & 54.53\% \\
9 & seq  & 41/35 & 94        & 1917                & 188827 & \cite{ws:mp} & 113       & 2239     & 84284 & 55.36\% \\
10 & t481 & 16/1  & 19        & 13                  & 220    & \cite{ws:mp} & 19        & 13       & 220   & 0\% \\
11 & vg2  & 25/8  & 51        & 207                 & 16525  & \cite{ws:mp} & 76        & 543      & 11709 & 29.14\% \\
12 & z4   & 7/4   & 14        & 36                  & 512    & \cite{ws:mp} & 20        & 78       & 484   & 5.47\% \\ \hline
13 & apex4 & 9/19 & 35        & 5131                & 228015 & \cite{mish}  & 61        & 5409     & 170541 & 25.21\% \\
14 & apla & 10/12 & 29        & 70                  & 3390   & \cite{mish}  & 40        & 244      & 1709   & 49.59\% \\
15 & bbm  & 4/4   & 10        & 16                  & 224    & \cite{mish}  & 17        & 42       & 164    & 26.79\% \\
16 & co14 & 14/1  & 26        & 14                  & 1610   & \cite{mish}  & 32        & 60       & 1070   & 33.54\% \\
17 & cordic & 23/2 & 40       & 1546                & 188715 & \cite{mish}  & 57        & 1686     & 127615 & 32.38\% \\
18 & cu   & 14/11 & 33        & 27                  & 1110   & \cite{mish}  & 39        & 93       & 631    & 43.15\% \\
19 & decod & 16/5 & 24        & 83                  & 1931   & \cite{mish}  & 41        & 193      & 847    & 56.14\% \\
20 & f51m & 14/8  & 34        & 369                 & 25155  & \cite{mish}  & 52        & 523      & 21953 & 12.73\% \\
21 & root & 8/5   & 19        & 67                  & 2605   & \cite{mish}  & 27        & 185      & 1786  & 31.44\% \\
22 & sqr6 & 6/12  & 22        & 59                  & 955    & \cite{mish}  & 30        & 109      & 655   & 31.41\% \\
23 & sqrt8 & 8/4  & 17        & 27                  & 495    & \cite{mish}  & 21        & 67       & 405   & 18.18\% \\
24 & table3 & 14/14 & 40      & 802                 & 74530  & \cite{mish}  & 63        & 1578     & 30320 & 59.32\% \\ \hline
25 & cycle10\_2 & 12/12 & 20  & 19               & 727    & \cite{www:rlsb} & 22        & 59       & 469   & 35.49\% \\
26 & cycle17\_3 & 20/20 & 35  & 48               & 3388   & \cite{www:rlsb} & 38        & 164      & 1824  & 46.16\% \\
27 & mod1024adder & 20/20 & 28 & 55              & 1435   & \cite{www:rlsb} & 30        & 139      & 1011  & 29.55\% \\
28 & \!mod1048576adder & 40/40 & 58 & 210        & 12090  & \cite{www:rlsb} & 59        & 588      & 6485  & 46.36\% \\
29 & nth\_prime6\_inc & 6/6 & 9 & 55             & 592    & \cite{www:rlsb} & 14        & 75       & 583   & 1.52\% \\ \hline
\end{tabular}
\label{tab:bench}
\end{center}
\end{footnotesize}
\end{table}

\section{Advantages and Limitations}

The control reduction algorithm we introduced in this paper has the following practical
advantages and limitations.
\begin{enumerate}
\item Advantages:
	\begin{itemize}
	\item Due to the structure of the circuits generated, this algorithm usually finds simplifications
	in the circuits generated by EXORCISM-4 \cite{ws:mp} or other ESOP synthesizers since every two gates commute,
	and MMD \cite{ar:mmd05} since it tends to use one control for a large number of sequential gates.
	\item This algorithm is particularly useful in compiling the Boolean {\tt if-then-else} type statement in a quantum
	programming language (previously mentioned in \cite{ar:spmh}). Indeed, the statement
	{\tt if} $x$ {\tt then} $A$ {\tt else} $B$ can be implemented such as shown in Figure \ref{fig:if}.
	The bottom circuit allows execution of statements $A$ and $B$ in parallel, which may be particularly helpful in the scenario
	when $A$, $B$, $AB^{-1}$ and $B^{-1}A$ have relatively high implementation costs, and a faster implementation is preferred.
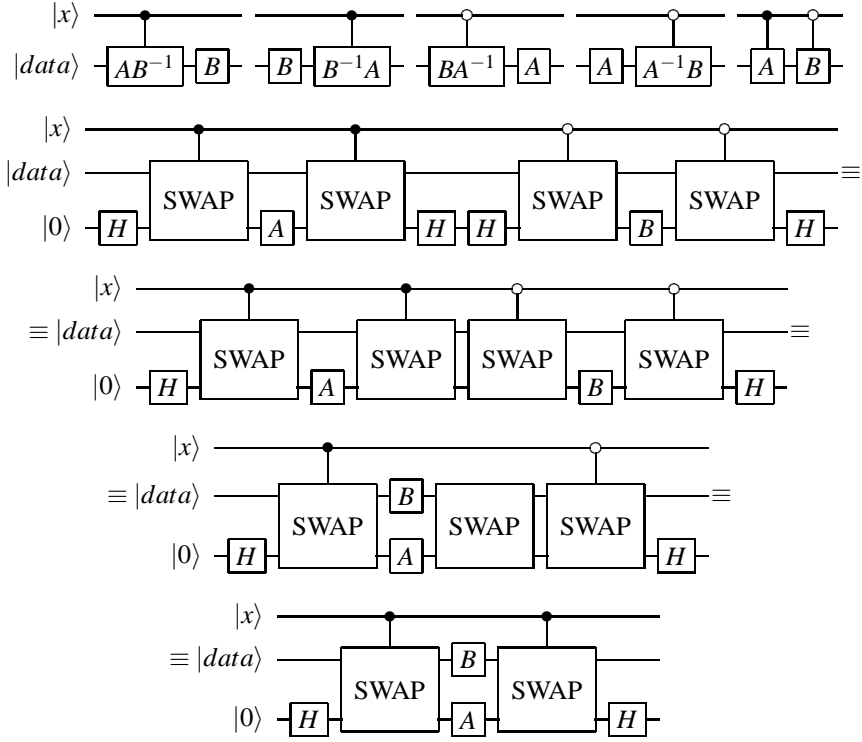
\begin{figure*}
\[
\Qcircuit @C=0.5em @R=1em {
\lstick{\ket{x}}    & \ctrl{1}       & \qw      & \qw & \;\; & \qw     & \ctrl{1}       & \qw & \;\; &
\ctrlo{1}      & \qw      & \qw & \;\; & \qw     & \ctrlo{1}      & \qw  & \;\; & \ctrl{1} & \ctrlo{1} & \qw\\
\lstick{\ket{data}} & \gate{AB^{-1}} & \gate{B} & \qw & \;\; & \gate{B}& \gate{B^{-1}A} & \qw & \;\; &
\gate{BA^{-1}} & \gate{A} & \qw & \;\; & \gate{A}& \gate{A^{-1}B} & \qw  & \;\; & \gate{A} & \gate{B}  & \qw\\
}
\]

\[
\Qcircuit @C=0.5em @R=1em {
\lstick{\ket{x}}    &\qw      & \ctrl{1}                   & \qw     & \ctrl{1}                   & \qw
                    &\qw      & \ctrlo{1}                  & \qw     & \ctrlo{1}                  & \qw     & \qw & \\
\lstick{\ket{data}} &\qw      & \multigate{1}{\text{SWAP}} & \qw     & \multigate{1}{\text{SWAP}} & \qw
                    &\qw      & \multigate{1}{\text{SWAP}} & \qw     & \multigate{1}{\text{SWAP}} & \qw     & \qw & \equiv \\
\lstick{\ket{0}}    &\gate{H} & \ghost{SWAP}               & \gate{A}& \ghost{SWAP}               & \gate{H}
                    &\gate{H} & \ghost{SWAP}               & \gate{B}& \ghost{SWAP}               & \gate{H}& \qw & \\
}
\]

\[
\Qcircuit @C=0.5em @R=1em {
\lstick{\ket{x}}          &\qw      & \ctrl{1}                   & \qw     & \ctrl{1}
                    & \ctrlo{1}                  & \qw     & \ctrlo{1}                  & \qw     & \qw & \\
\lstick{\equiv\ket{data}} &\qw      & \multigate{1}{\text{SWAP}} & \qw     & \multigate{1}{\text{SWAP}}
                    & \multigate{1}{\text{SWAP}} & \qw     & \multigate{1}{\text{SWAP}} & \qw     & \qw & \equiv \\
\lstick{\ket{0}}          &\gate{H} & \ghost{SWAP}               & \gate{A}& \ghost{SWAP}
                    & \ghost{SWAP}               & \gate{B}& \ghost{SWAP}               & \gate{H}& \qw & \\
}
\]

\[
\Qcircuit @C=0.5em @R=1em {
\lstick{\ket{x}}          &\qw      & \ctrl{1}                   & \qw     & \qw         & \ctrlo{1}    & \qw     & \qw & \\
\lstick{\equiv\ket{data}} &\qw      & \multigate{1}{\text{SWAP}} & \gate{B}& \multigate{1}{\text{SWAP}}
	& \multigate{1}{\text{SWAP}} & \qw     & \qw & \equiv \\
\lstick{\ket{0}}          &\gate{H} & \ghost{SWAP}               & \gate{A}& \ghost{SWAP}& \ghost{SWAP} & \gate{H}& \qw & \\
}
\]

\[
\Qcircuit @C=0.5em @R=1em {
\lstick{\ket{x}}          &\qw      & \ctrl{1}                   & \qw     & \ctrl{1}                   & \qw     & \qw \\
\lstick{\equiv\ket{data}} &\qw      & \multigate{1}{\text{SWAP}} & \gate{B}& \multigate{1}{\text{SWAP}} & \qw     & \qw \\
\lstick{\ket{0}}          &\gate{H} & \ghost{SWAP}               & \gate{A}& \ghost{SWAP}               & \gate{H}& \qw \\
}
\]
\centering
\caption{Implementation of the {\tt if} $x$ {\tt then} $A$ {\tt else} $B$ statement. Top: five basic ways to implement
this statement.  Circuit equivalence from Figure \ref{fig:main} may be applied to simplify each of these five implementations.
Middle to bottom: an illustration of how the circuit equivalence from Figure \ref{fig:main} helps to
simplify the top right implementation.}
\label{fig:if}
\end{figure*}

	\end{itemize}
\item Limitations:
	\begin{itemize}
	\item This algorithm is unlikely to find simplification in circuits dominated by small gates, {\it e.g.}, single- and two-qubit gates,
	such as those generated by \cite{ar:mmd,ar:szss,ar:spmh}.
	\item A sufficient number of auxiliary qubits set to value $\ket{0}$ needs to be made available for the algorithm to work efficiently.
	However, the performance improves as the number of auxiliary qubits carrying value $\ket{0}$ grows (for example, we did not test nested 
	application of our algorithm, but expect the results may improve compared to those reported in this paper).
	\end{itemize}
\end{enumerate}

\begin{table*}
\scriptsize
\caption{The distribution of the number of gates for the circuits reported in Table \ref{tab:bench}.}
\begin{tabular}{|l|c|c|c|c|c|c|c|c|c|c|c|c|c|c|c|c|c|c|c|c|c|c|c|}
\hline
\!\!ckt\#\!\!\!\!&\!\!T1\!\!&\!\!T2\!\!&\!\!T3\!\!&\!\!T4\!\!&\!\!T5\!\!&\!\!T6\!\!&\!\!T7\!\!&\!\!T8\!\!&\!\!T9\!\!&\!\!T10\!\!&\!\!T11\!\!&
\!\!T12\!\!&\!\!T13\!\!&\!\!T14\!\!&\!\!T15\!\!&\!\!T16\!\!&\!\!T17\!\!&\!\!T18\!\!&\!\!T19\!\!&\!\!T20\!\!&\!\!T21\!\!&\!\!F3\!\!&\!\!H\!\!\\
\hline
\hline
1b &0&\!17\!&7&12&10&4&5&6&0&0&0&0&0&0&0&0&0&0&0&0&0&0&0	\\
1a &0&\!27\!&20&10&4&7&1&0&0&0&0&0&0&0&0&0&0&0&0&0&0&52&\!20\!\\
\hline
2b &6     &0     &23&18&29&35&45&32&0&0&0&0&0&0&0&0&0&0&0&0&0&0&0\\
2a &\!19\!&\!16\!&47&37&45&32&0&0&0&0&0&0&0&0&0&0&0&0&0&0&0&\!100\!&\!34\!\\
\hline
3b &5&0&6&4&9&4&9&6&0&0&0&0&0&0&0&0&0&0&0&0&0&0&0\\
3a &8&1&13&11&11&2&1&0&0&0&0&0&0&0&0&0&0&0&0&0&0&42&\!24\!\\
\hline
4b &0&2&10&2&7&14&33&28&16&8&0&0&0&0&0&0&0&0&0&0&0&0&0\\
4a &0&9&39&38&31&12&2&3&0&0&0&0&0&0&0&0&0&0&0&0&0&\!140\!&\!22\!	\\
\hline
5b &4&3&5&4&3&0&3&5&13&10&9&2&2&4&1&0&0&2&0&0&0&0&0\\
5a &9&7&8&11&15&13&3&2&5&1&0&0&2&0&0&0&0&0&0&0&0&78&\!36\!\\
\hline
6b &0&0&0&0&3&9&15&10&11&2&0&0&0&0&0&0&0&0&0&0&0&0&0	\\
6a &0&0&10&17&11&9&12&1&0&0&0&0&0&0&0&0&0&0&0&0&0&76&\!16\!\\
\hline
7b &0&1&0&3&0&6&0&9&0&9&0&7&0&4&0&1&0&0&0&0&0&0&0\\
7a &3&2&7&7&7&10&1&6&0&4&0&1&0&0&0&0&0&0&0&0&0&60&\!26\!\\
\hline
8b &0&2&1&0&0&0&0&3&17&28&7&0&0&0&0&0&0&0&0&0&0&0&0\\
8a &0&6&6&30&9&10&4&2&1&0&0&0&0&0&0&0&0&0&0&0&0&78&\!18\!\\
\hline
9b &0     &2     &14&1&0&8&2&33&\!128\!&\!187\!&\!115\!&\!126\!&\!189\!&\!151\!&\!209\!&\!245\!&\!198\!&\!141\!&76&52&40&0&0\\
9a &\!19\!&\!69\!&\!204\!&\!194\!&\!237\!&\!213\!&\!168\!&\!283\!&\!184\!&\!171\!&\!100\!&53&40&0&0&0&0&0&0&0&0&\!252\!&\!52\!\\
\hline
10b &1&0&4&0&8&0&0&0&0&0&0&0&0&0&0&0&0&0&0&0&0&0&0\\
10a &1&0&4&0&8&0&0&0&0&0&0&0&0&0&0&0&0&0&0&0&0&0&0\\
\hline
11b &0&0     &4&7&16&12&10&16&8&24&30&26&16&16&4&4&0&4&6&0&4&0&0\\
11a &0&\!14\!&33&28&12&20&10&28&26&18&18&4&4&0&0&6&0&4&0&0&0&\!264\!&\!54\!\\
\hline
12b &0&2&14&10&6&4&0&0&0&0&0&0&0&0&0&0&0&0&0&0&0&0&0\\
12a &0&4&26&10&0&2&0&0&0&0&0&0&0&0&0&0&0&0&0&0&0&26&\!10\!\\
\hline
\hline
13b &0&0&0&\!171\!&\!537\!&\!\!1188\!\!&\!\!1460\!\!&\!\!1167\!\!&\!504\!&\!104\!&0&0&0&0&0&0&0&0&0&0&0&0&0\\
13a &8&1&\!165\!&\!651\!&\!\!1265\!\!&\!\!1520\!\!&\!\!1072\!\!&\!367\!&\!86\!&0&0&0&0&0&0&0&0&0&0&0&0&\!222\!&\!52\!\\
\hline
14b &0&0     &0&6&5&7&16&18&13&5&0&0&0&0&0&0&0&0&0&0&0&0&0\\
14a &3&\!24\!&25&11&12&3&2&4&0&0&0&0&0&0&0&0&0&0&0&0&0&\!136\!&\!24\!	\\
\hline
15b &0&4&4&0&8&0&0&0&0&0&0&0&0&0&0&0&0&0&0&0&0&0&0\\
15a &2&4&10&0&2&0&0&0&0&0&0&0&0&0&0&0&0&0&0&0&0&12&\!12\!	\\
\hline
16b &0&0&0&0&0&0&0&0&0&0&0&0&0&14&0&0&0&0&0&0&0&0&0\\
16a &0&0&0&12&0&0&0&0&0&0&6&0&0&2&0&0&0&0&0&0&0&30&\!10\!\\
\hline
17b &1&0&0&0&0&5&0&0&0&0&0&388&0&0&768&0&384&0&0&0&0&0&0\\
17a &1&0&3&2&4&2&0&388&0&0&768&0&384&0&0&0&0&0&0&0&0&92&\!42\!\\
\hline
18b &1&0&2&0&7&4&5&4&0&0&4&0&0&0&0&0&0&0&0&0&0&0&0\\
18a &8&6&6&0&9&2&2&0&0&0&0&0&0&0&0&0&0&0&0&0&0&42&\!18\!\\
\hline
19b &0&1     &8&12&46&16&0&0&0&0&0&0&0&0&0&0&0&0&0&0&0&0&0\\
19a &8&\!17\!&\!50\!&12&2&0&0&0&0&0&0&0&0&0&0&0&0&0&0&0&0&70&\!34\!\\
\hline
20b &0&0&9&16&22&27&26&23&47&46&60&59&23&6&5&0&0&0&0&0&0&0&0\\
20a &4&8&\!29\!&25&21&27&21&47&46&60&59&23&6&5&0&0&0&0&0&0&0&\!106\!&\!36\!\\
\hline
21b &0&5&6&2&11&6&11&13&13&0&0&0&0&0&0&0&0&0&0&0&0&0&0\\
21a &1&6&\!22\!&20&13&16&1&0&0&0&0&0&0&0&0&0&0&0&0&0&0&88&\!18\!\\
\hline
22b &0&5&\!24\!&12&1&14&3&0&0&0&0&0&0&0&0&0&0&0&0&0&0&0&0\\
22a &2&5&\!37\!&17&1&1&0&0&0&0&0&0&0&0&0&0&0&0&0&0&0&30&\!16\!\\
\hline
23b &0&5&7&4&4&3&3&1&0&0&0&0&0&0&0&0&0&0&0&0&0&0&0\\
23a &0&5&\!13\!&11&2&0&0&0&0&0&0&0&0&0&0&0&0&0&0&0&0&24&\!12\!\\
\hline
24b &0&0     &0&0&6&10&0&10&61&89&\!154\!&\!169\!&\!154\!&\!110\!&39&0&0&0&0&0&0&0&0\\
24a &6&\!55\!&73&\!120\!&\!189\!&\!143\!&\!105\!&73&24&36&16&4&2&0&0&0&0&0&0&0&0&\!684\!&\!48\!\\
\hline
\hline
25b &0&2&2&2&2&2&2&2&2&2&1&0&0&0&0&0&0&0&0&0&0&0&0\\
25a &2&4&4&4&5&4&0&0&0&0&0&0&0&0&0&0&0&0&0&0&0&24&\!12\!\\
\hline
26b &0&3&3&3&3&3&3&3&3&3&3&3&3&3&3&3&2&1&0&0&0&0&0\\
26a &6&9&9&12&5&3&3&5&7&1&0&0&0&0&0&0&0&0&0&0&0&84&\!20\!\\
\hline
27b &0&\!10\!&9&8&7&6&5&4&3&2&1&0&0&0&0&0&0&0&0&0&0&0&0\\
27a &6&\!16\!&16&15&10&4&0&0&0&0&0&0&0&0&0&0&0&0&0&0&0&60&\!12\!\\
\hline
28b &0     &\!20\!&19&18&17&16&15&14&13&12&11&10&9&8&7&6&5&4&3&2&1&0&0\\
28a &\!25\!&\!45\!&47&47&35&20&15&8&10&6&2&0&0&0&0&0&0&0&0&0&0&\!308\!&\!20\!\\
\hline
29b &5&\!12\!&14&11&11&2&0&0&0&0&0&0&0&0&0&0&0&0&0&0&0&0&0\\
29a &6&\!13\!&15&9&11&1&0&0&0&0&0&0&0&0&0&0&0&0&0&0&0&10&\!10\!\\
\hline
\end{tabular}
\label{tab:bench2}
\end{table*}

\section{Conclusions} \label{s:cfr}
In this paper, we presented an approach for systematic optimization of reversible circuits that trades in qubits to
achieve a lower implementation cost.  This may be of particular interest in practice when a multistage quantum
algorithm (including computations in the Boolean domain) needs to be executed on a quantum processor, and there are a number
of scrap qubits available to be used to optimize intermediate computations.  The proposed approach may be
extended to optimize quantum controlled transformations.

\section{Acknowledgments}

This article was based on work partially supported by the National Science Foundation, during
D. Maslov's assignment at the Foundation.

\end{document}